\documentclass[sn-mathphys-num]{sn-jnl}

\usepackage{graphicx}%
\usepackage{multirow}%
\usepackage{amsmath,amssymb,amsfonts}%
\usepackage{amsthm}%
\usepackage{mathrsfs}%
\usepackage[title]{appendix}%
\usepackage{xcolor}%
\usepackage{textcomp}%
\usepackage{manyfoot}%
\usepackage{booktabs}%
\usepackage{algpseudocode}%
\usepackage{listings}%
\usepackage{amsfonts}
\usepackage{bm}
\usepackage{braket}
\usepackage{float} 
\usepackage[ruled,vlined]{algorithm2e} 
\usepackage[utf8]{inputenc}   
\usepackage[T1]{fontenc}      
\usepackage{bbold}
\usepackage{float}

\usepackage[ruled,vlined]{algorithm2e} 
\usepackage{amssymb} 
\usepackage{bbold}   

\theoremstyle{thmstyleone}%
\theoremstyle{thmstyletwo}%

\theoremstyle{thmstylethree}%

\begin{document}

\title[Article Title]{Modified Conjugate Quantum Natural Gradient}


\author[1]{\fnm{Mourad} \sur{Halla}}


\affil[1]{\orgname{Deutsches Elektronen-Synchrotron DESY}, \orgaddress{\street{Platanenallee 6}, \postcode{15738} \city{Zeuthen}, \country{Germany}}}

\abstract{
The efficient optimization of variational quantum algorithms (VQAs) is critical for their successful application in quantum computing. The Quantum Natural Gradient (QNG) method, which leverages the geometry of quantum state space, has demonstrated improved convergence compared to standard gradient descent [Quantum \textbf{4}, 269 (2020)]. In this work, we introduce the Modified Conjugate Quantum Natural Gradient (CQNG), an optimization algorithm that integrates QNG with principles from the nonlinear conjugate-gradient method. Unlike QNG, which employs a fixed learning rate, CQNG dynamically adjusts its hyperparameters at each step, enhancing both efficiency and flexibility. Numerical simulations show that CQNG achieves faster convergence and reduces quantum-resource requirements compared to QNG across various optimization scenarios, even when strict conjugacy conditions are not fully satisfied—hence the term “Modified Conjugate.” These results highlight CQNG as a promising optimization technique for improving the performance of VQAs.
}





\maketitle

\section{Introduction}

Quantum computing has emerged as a powerful paradigm for solving computational problems that surpass the capabilities of classical algorithms. One of the most promising approaches in this domain is \textit{Variational Quantum Algorithms (VQAs)} \cite{Cerezo2021, McClean2016, Bharti2022}, which have been widely studied for applications in quantum chemistry, materials science, and condensed matter physics. Among them, the \textit{Variational Quantum Eigensolver (VQE)} \cite{Peruzzo2014} plays a crucial role in approximating ground-state energies of quantum systems, offering a hybrid quantum-classical optimization framework.

VQE leverages the flexibility of parameterized quantum circuits (ansatz), which are iteratively optimized to minimize an energy-based cost function. The optimization is performed using suitable optimizers, while quantum processors handle state preparation and measurement. This hybrid approach enables VQE to efficiently explore the solution space, making it particularly relevant for near-term noisy intermediate-scale quantum (NISQ) devices.

Optimization strategies are crucial for the efficiency of VQAs, as they determine both the convergence rate and the quality of the approximated solutions. Traditional methods, such as gradient descent (GD), are often insufficient to handle the complex characteristics of quantum optimization landscapes. These landscapes, characterized by non-convexity, noise, and barren plateaus \cite{McClean2018}, require more advanced algorithms to accelerate convergence and enhance robustness. Moreover, each optimization step in a VQA typically demands a large number of circuit measurements, which becomes a serious overhead as system size grows. It is therefore important to develop algorithms that reduce the number of measurements per iteration, as demonstrated in quantum chemistry in \cite{Patel2025} and \cite{Gonthier2022}.

One of the sophisticated optimizers, as an alternative to GD, is the Quantum Natural Gradient (QNG) algorithm \cite{Stokes2020}, which leverages the geometry of the parameter space to enhance optimization in variational quantum algorithms. By utilizing the Fubini-Study metric or quantum Fisher information \cite{Meyer}, QNG captures the local curvature of the quantum state manifold, aligning updates with the natural geometry to accelerate convergence and identify regions of high curvature early in the optimization process \cite{Katabarwa2022}. Extensions to QNG have improved its robustness, including adaptations for noisy and nonunitary circuits \cite{Koczor}, stochastic techniques for approximating the quantum Fisher information matrix \cite{Gacon}, the Quantum Random Natural Gradient \cite{Kolotouros}, and the Quantum Natural Gradient with Geodesic Correction \cite{Halla}, the quantum Broyden adaptive natural gradient (qBang) \cite{qBang} making it a powerful optimizer for VQAs. In this work, we further extend QNG by integrating the nonlinear conjugate gradient method.

The \textit{nonlinear conjugate gradient method} is a widely used optimization technique for tackling high-dimensional \textit{nonlinear optimization problems}~\cite{Hestenes1952,Fletcher1964}. At each iteration, it constructs new search directions by conjugating the residuals from previous steps, effectively forming a \textit{Krylov subspace}~\cite{Saad2003}. This method is computationally efficient, requiring only the current gradient and the previous search direction, thereby reducing memory complexity. Building on these foundations, the \textit{classical natural gradient} method~\cite{Amari1998} has been integrated with nonlinear conjugate gradient techniques by replacing the Euclidean gradient with the classical natural gradient~\cite{Pascanu2013}. Inspired by this approach, we extend these principles to the quantum domain, applying the nonlinear conjugate gradient method to the QNG framework. This integration aims to enhance convergence efficiency in variational quantum algorithms.

This manuscript is structured as follows: Section~\ref{Theoty:QNG} provides an overview of QNG and VQE. In Section~\ref{Theoty:CQNG}, we introduce CQNG as an extension of QNG. Section~\ref{Experimental} presents numerical simulations demonstrating that CQNG accelerates convergence and reduces quantum-resource requirements. Finally, Section~\ref{outlook} summarizes the key findings and discusses potential future extensions.

\section{Theory}
\subsection{\textbf{Quantum Natural Gradient and VQE}}
\label{Theoty:QNG}

This section provides an overview of the VQE and its optimization strategies, focusing on noise-free variational quantum circuits. For a comprehensive review of VQAs, readers are referred to \cite{Cerezo2021,McClean2016,Bharti2022}. An in-depth discussion of QNG can be found in \cite{Stokes2020}.

A typical circuit in VQAs is constructed as a sequence of unitary operations:
\begin{equation}
U_L(\bm{\theta}) = V_L(\bm{\theta}_L) W_L \cdots V_1(\bm{\theta}_1) W_1,
\end{equation}
where \(V_l(\bm{\theta}_l)\) are parameterized unitary operators, \(W_l\) are fixed unitaries, and \(\bm{\theta} = (\bm{\theta}_1 \oplus \cdots \oplus \bm{\theta}_L) \in \mathbb{R}^d\) represents the concatenated parameter vector.

The objective of VQE is to find the ground state of the observable \(\hat{O}\) by minimizing a cost function, typically defined as the expectation value of \(\hat{O}\) with respect to the parameterized quantum state \(|\psi(\bm{\theta})\rangle = U(\bm{\theta})|\psi_0\rangle\):
\begin{equation}
\mathcal{L}(\bm{\theta}) = \bra{\psi(\bm{\theta})} \hat{O} \ket{\psi(\bm{\theta})},
\end{equation}
where \(|\psi_0\rangle\) is the initial state. The minimization of \(\mathcal{L}(\bm{\theta})\) over the parameter space \(\bm{\theta}\) yields an approximation of the ground state energy and its corresponding eigenstate. Gradient descent achieves this by iteratively updating the parameters:
\begin{equation}
\bm{\theta}_{t+1} = \bm{\theta}_t - \eta\, \boldsymbol{\nabla} \mathcal{L}(\bm{\theta}_t),
\label{GD}
\end{equation}
where $
\boldsymbol{\nabla} := (\partial_1, \dots, \partial_l) = \left( \frac{\partial}{\partial \theta_1}, \dots, \frac{\partial}{\partial \theta_l} \right)
$ denotes the gradient operator with respect to the parameter vector $
\bm{\theta} = (\theta_1, \dots, \theta_l)$, \(\eta > 0\) is the learning rate, and \(t\) denotes the iteration step. The gradient of the cost function \( \boldsymbol{\nabla} \mathcal{L}(\bm{\theta}_t) \) can be calculated using the parameter-shift rule \cite{Schuld,Mari,Wierichs}.

In standard Gradient Descent (\ref{GD}), parameter updates are performed in a flat Euclidean space, which does not accurately capture the curved geometry of the quantum state manifold. QNG addresses this limitation by incorporating the Riemannian gradient, \( \bm{F}^{-1} \boldsymbol{\nabla} \mathcal{L}(\bm{\theta}_t) \), into the update rule:

\begin{equation}
\bm{\theta}_{t+1} = \bm{\theta}_t - \eta \bm{F}^{-1} \boldsymbol{\nabla} \mathcal{L}(\bm{\theta}_t),
\label{eq:QNG}
\end{equation}
where the Fubini–Study metric \( \bm{F} \) (with \( \bm{F}^{-1} \) as its inverse) is defined component-wise as:
\begin{equation}
F_{ij} = \mathrm{Re} \left( \left\langle \partial_i \psi \middle| \partial_j \psi \right\rangle \right) - \left\langle \partial_i \psi \middle| \psi \right\rangle \left\langle \psi \middle| \partial_j \psi \right\rangle.
\label{Fubini}
\end{equation}
The QNG is a robust and effective optimizer; however, it incurs additional computational overhead due to the calculation of the metric. Specifically, evaluating the full Fubini-Study metric \( F_{ij} \) for a circuit with \( m \) parameters requires \( O(m^2) \) function evaluations, making it computationally demanding for large-scale systems. To mitigate this, the block-diagonal approximation is employed, significantly reducing the complexity. The key idea involves defining subcircuits between layers \( l_1 \) and \( l_2 \) as:

\begin{equation}
U_{[l_1:l_2]} = V_{l_2} W_{l_2} \cdots V_{l_1} W_{l_1},
\end{equation}
and expressing the full circuit recursively as:
\begin{equation}
U_L(\bm{\theta}) = U_{(l:L]} V_l W_l U_{[1:l)},
\end{equation}
where \((l:L] = [l-1:L]\) and \([1:l) = [1:l-1]\). The quantum state at the \(l\)-th layer is defined as:
\begin{equation}
\psi_l := U_{[1:l]}|0\rangle.
\end{equation}
For each layer \(l \in [L]\), Hermitian generator matrices \(K_i\) and \(K_j\) are defined such that:
\begin{equation}
\partial_i V_l(\bm{\theta}_l) = -i K_i V_l(\bm{\theta}_l),
\end{equation}
\begin{equation}
\partial_j V_l(\bm{\theta}_l) = -i K_j V_l(\bm{\theta}_l),
\end{equation}
where \( [K_i, K_j] = 0 \) for distinct parameters \(i \neq j\). The block-diagonal approximation of the Fubini-Study metric \(F_{ij}\) for layer \(l\) is then given by:
\begin{equation}
F_{ij}^{(l)} = \langle \psi_l | K_i K_j | \psi_l \rangle - \langle \psi_l | K_i | \psi_l \rangle \langle \psi_l | K_j | \psi_l \rangle.
\end{equation}
If we consider circuits composed of single-qubit Pauli rotations, expressed as:
\begin{equation}
V_l(\bm{\theta}_l) = \bigotimes_{k=1}^n R_{P_{l,k}}(\bm{\theta}_{l,k}), \quad R_{P_{l,k}}(\bm{\theta}_{l,k}) = \exp\left(-i \frac{\bm{\theta}_{l,k}}{2} P_{l,k}\right),
\end{equation}
where \( P_{l,k} \in \{\sigma_x, \sigma_y, \sigma_z\} \) are Pauli matrices acting on the \( k \)-th qubit. The Hermitian generator for parameter \( \bm{\theta}_{l,k} \) is:
\begin{equation}
K_i = \frac{1}{2} \mathbb{1}^{[1,i)} \otimes P_{l,i} \otimes \mathbb{1}^{(i,n]},
\end{equation}
where \( \mathbb{1}^{[1,i)} = \bigotimes_{1 \leq j < i} \mathbb{1} \) is the identity on preceding qubits. These generators satisfy \( [K_i, K_j] = 0 \) and \( P_{l,i}^2 = \mathbb{1} \).

The metric block for layer \(l\) requires the evaluation of the quantum expectation value \( \bra{\psi_l} \hat{A} \ket{\psi_l} \), where \( \hat{A} \) is an operator from:
\begin{equation}
S_l = \{P_{l,i} \mid 1 \leq i \leq n\} \cup \{P_{l,i}P_{l,j} \mid 1 \leq i < j \leq n\}.
\end{equation}
Since all operators in \( S_l \) commute, a single quantum measurement per layer is sufficient, reducing the required number of state preparations from \( |S_l| = n(n+1)/2 \).

To further improve optimization, in this work we propose integrating QNG with the nonlinear conjugate gradient method, which is discussed in the following section.

\subsection{\textbf{Modified Conjugate Quantum Natural Gradient}}
\label{Theoty:CQNG}
Quantum Natural gradient descent considers optimization as moving in the steepest direction along a curved parameter space, yet it remains fundamentally a first-order method. Directly including second-order information of the cost function, such as the Hessian, to improve convergence is often computationally prohibitive. A viable alternative that avoids explicitly computing the Hessian is the Nonlinear Conjugate Gradient method, which only needs to combine the new natural gradient with the previous search direction, without calculating second-order derivatives explicitly. In this section, we enhance the QNG algorithm by incorporating the nonlinear conjugate gradient method. For a review of conjugate gradient methods, we recommend the book \cite{Andrei2020}.

The Modified Conjugate Quantum Natural Gradient (CQNG) extends the QNG algorithm (\ref{eq:QNG}) by incorporating principles from the nonlinear conjugate gradient method. The parameter update rule for CQNG, for steps \( t \geq 0 \), is defined as:
\begin{equation}
\bm{\theta}_{t+1} = \bm{\theta}_t + \alpha_t \bm{d}_t,
\label{eq:update_rule}
\end{equation}
where the search directions \( \bm{d}_t \in \mathbb{R}^n \) are computed as:
\begin{equation}
\bm{d}_{t} =
\left\{
\begin{array}{ll}
    -\bm{F}^{-1} \boldsymbol{\nabla} \mathcal{L}(\bm{\theta}_t) & \mathrm{if} \ t = 0, \\
    -\bm{F}^{-1} \boldsymbol{\nabla} \mathcal{L}(\bm{\theta}_t) + \beta_t \bm{d}_{t-1} & \mathrm{if} \ t \geq 1.
\end{array}
\right.
\label{eq:search_direction}
\end{equation} 
Here, \( \alpha_t \in \mathbb{R}^+ \) is the step size, \( \bm{F}^{-1} \) denotes the inverse of the Fubini–Study metric, and \( \beta_t \in \mathbb{R} \) is the conjugate coefficient that incorporates the influence of the previous search direction \( \bm{d}_{t-1} \) and ensures that the conjugacy condition with respect to the Hessian \( \bm{H} \) is satisfied.

\begin{equation}
\bm{d}_t^T \bm{H} \bm{d}_{t-1} = 0.
\end{equation}
However, unlike standard conjugate gradient methods that do not account for manifold geometry, the natural gradient approach defines search directions $\bm{d}_t$ and $\bm{d}_{t-1}$ using the Fubini-Study metric. This creates distinct local geometries around each step: the geometry at the previous direction $\bm{d}_{t-1}$ is defined by $\bm{F}_{t-1}$, while the geometry at the new direction $\bm{d}_t$ is defined by $\bm{F}_t$, where in general $\bm{F}_{t-1} \neq \bm{F}_t$.

Therefore, it is necessary to first transport $\bm{d}_{t-1}$ into the space of the natural gradient, specifically into $\bm{F}_t^{-1} \boldsymbol{\nabla} \mathcal{L}$, before applying standard formulas for $\beta_t$, such as Polak–Ribi\`{e}re~\cite{Polak1969}, Fletcher–Reeves~\cite{Fletcher1964}, or Dai–Yuan~\cite{Dai1999}. Moreover, directly applying these standard formulas is no longer straightforward, since the line search must follow the geodesic of the manifold. Together, these issues make enforcing the conjugacy condition both difficult and computationally expensive in practice for VQAs.

To address this bottleneck, we adopt a more practical strategy inspired by techniques from classical deep networks in~\cite{Pascanu2013}. Rather than using a fixed exact formula for $\beta_t$, we compute both $\alpha_t$ and $\beta_t$ simultaneously at each iteration by solving the following optimization problem:

\begin{equation}
(\alpha_t, \beta_t) = \arg\min_{\alpha, \beta} \mathcal{L}\left( \bm{\theta}_t + \alpha_t\; \bm{F}^{-1} \boldsymbol{\nabla}  \mathcal{L}(\bm{\theta}_t) + \beta_t\; \bm{d}_{t-1} \right).
\label{eq:update_beta}
\end{equation}
By expanding $\mathcal{L}$ around $\bm{\theta}_{t}$ using a second-order Taylor approximation and applying the stationarity condition, it can be shown that this method yields a search direction that is conjugate to the previous one in the Euclidean setting. Since we do not perform a multidimensional line search along geodesics or transport the previous direction $\bm{d}_{t-1}$ into the current tangent space, the resulting update is only an approximation of the true locally conjugate direction on the manifold.

This adaptive strategy enhances flexibility and robustness by jointly optimizing the step size and conjugate coefficient at each iteration, even when strict conjugacy conditions are not always satisfied. Due to this modification, we refer to the approach as the "Modified Conjugate". The main steps of the CQNG algorithm are summarized in Algorithm~\ref{alg:CQNG}. In the next section, we present numerical simulations demonstrating its improved convergence performance compared to standard QNG.

\vspace{0.5cm}

\small
\begin{algorithm}[H]
\caption{Modified Conjugate Quantum Natural Gradient (CQNG)}
\label{alg:CQNG}
\KwIn{Problem Hamiltonian $\hat{H}$, Ansatz $\ket{\psi(\bm{\theta})}$, initial parameters $\bm{\theta}_0$, objective function $\mathcal{L}(\bm{\theta})$, maximum iterations $T$, regularization parameter $\lambda$, $\alpha_0 \in \mathbb{R}^+$, $\beta_0 \in \mathbb{R}$.}
\KwOut{Optimized parameters $\bm{\theta}_T$.}

\For{$t = 0$ \textbf{to} $T-1$}{
    Compute the gradient: $\boldsymbol{\nabla} \mathcal{L}(\bm{\theta}_t)$\;
    Compute and regularize the metric tensor (\( \bm{I} \): identity matrix): $\bm{F} \leftarrow \bm{F}(\bm{\theta}_t) + \lambda \bm{I} $\;
    Compute the inverse metric tensor: $\bm{F}^{-1}$\;

    \eIf{$t = 0$}{
        Set the search direction: $\bm{d}_t \leftarrow -\bm{F}^{-1} \boldsymbol{\nabla} \mathcal{L}(\bm{\theta}_t)$\;
    }{
        Solve for optimal $\alpha_t$ and $\beta_t$ using an optimizer with initial values $\alpha_0$ and $\beta_0$:
        \begin{equation*}
        (\alpha_t, \beta_t) = \arg\min_{\alpha, \beta} \mathcal{L}\left( \bm{\theta}_t + \alpha \bm{d}_t + \beta \bm{d}_{t-1} \right).
        \end{equation*}
        
        \If{optimization fails}{
            Set default values: $\alpha_t \leftarrow \alpha_0$, $\beta_t \leftarrow 0$\;
        }
        
        Update the search direction:
        \begin{equation*}
        \bm{d}_t \leftarrow -\bm{F}^{-1} \boldsymbol{\nabla} \mathcal{L}(\bm{\theta}_t) + \beta_t \bm{d}_{t-1}.
        \end{equation*}
    }

    Update parameters:
    \begin{equation*}
    \bm{\theta}_{t+1} \leftarrow \bm{\theta}_t + \alpha_t \bm{d}_t.
    \end{equation*}
}
\Return $\bm{\theta}_T$\;
\end{algorithm}

\section{Experimental Simulation}
\label{Experimental}

In this section, we present practical examples to demonstrate the accuracy and effectiveness of the conjugate method. For the simulations, we employ three distinct approaches. Example 1, as a pedagogical case, consists of a two-qubit simulation, building on analytical results from our previous work~\cite{Halla}. In Examples 2 and 3, we explore the Heisenberg model and molecular Hamiltonians—hydrogen four (H\textsubscript{4}), water (H\textsubscript{2}O), and carbon diatomic (C\textsubscript{2})—using the open-source software \texttt{PennyLane}~\cite{PennyLane}, illustrating the scalability of our method. We benchmark standard gradient descent (GD), both block-diagonal and full-metric variants of the quantum natural gradient (QNG (block-diag) and QNG), and both block-diagonal and full-metric variants of the conjugate quantum natural gradient (CQNG (block-diag) and CQNG).

\subsection{\textbf{Example 1: Two-Qubit Simulation}}
\label{Example1}

In this example, we determine the ground state energy (in Hartree) of the hydrogen molecule (\(H_2\)) using a two-qubit VQE. The system is described by the following Hamiltonian:
\begin{equation}
H = h (\sigma_z \otimes I + I \otimes \sigma_z) + J \sigma_x \otimes \sigma_x,
\end{equation}
where \(h = 0.4\) and \(J = 0.2\). Here, \(\sigma_z\) and \(\sigma_x\) are Pauli matrices acting on the qubits, with the parameters \(J\) and \(h\) controlling the interaction and magnetic field terms, respectively.

To approximate the ground state, we utilize a parameterized VQE ansatz designed to capture the system's entanglement and structure. This ansatz is given by:
\begin{equation}
|\psi\rangle = (CRY(\theta)_{q_0, q_1}) (CRX(\theta)_{q_0, q_1}) \left( R_y(2\theta_0) \otimes R_y(2\theta_1) \right) \ket{0} \otimes \ket{0}\\
\label{ansatz_ex2}
\end{equation}
where \(R_y(\theta) = e^{-i\theta\sigma_y/2}\) is a single-qubit rotation gate, and \(CRX, CRY\) are controlled rotation gates entangling the qubits.

The Fubini-Study metric, which governs the geometry of the parameter space, is computed as:
\begin{equation}
F =
\left(
\begin{array}{ccc}
1 & 0 & \cos(\theta_1) \sin(\theta_1) \\
0 & 1 & -\cos(\theta_0) \sin(\theta_0) \\
\cos(\theta_1) \sin(\theta_1) & -\cos(\theta_0) \sin(\theta_0) & \frac{1}{2} \left( 1 - \cos(2\theta_0) \cos(2\theta_1) \right)
\end{array}
\right).
\end{equation}

To prevent singularities in \(F\) during parameter updates, a regularization term \(\lambda\) is added to its diagonal, ensuring numerical stability when computing the inverse metric, \(F^{-1}\). The derivation of the inverse metric and additional details about this system are provided in \cite{Halla}.

\begin{figure}[H]
    \centering
    \includegraphics[width=\textwidth]{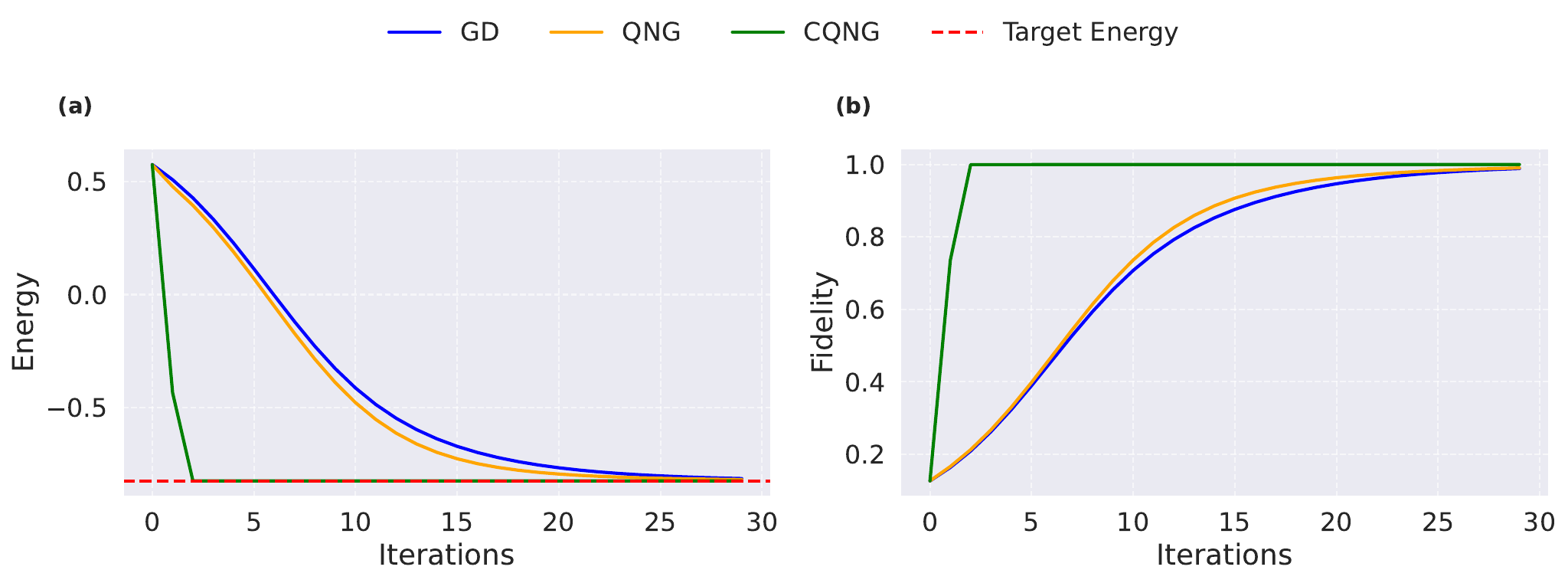}
   \caption{
    Energy convergence and fidelity to the ground state as a function of training iterations for GD, QNG, and CQNG optimizers. The initial parameters are set to $[-0.2,-0.2,0]$, and a learning rate $\eta = 0.05$ is used for GD and QNG, and CQNG. The initial values $\beta_0 = 0.1$ and $\alpha_0 = 0.05$ are used for the \texttt{COBYLA} optimizer, which dynamically optimizes $\alpha_t$ and $\beta_t$ at each step according to Eq.~(\ref{eq:update_beta}).}
    \label{fig:ex1_energy}
\end{figure}

\begin{figure}[H]  
    \centering
    \includegraphics[width=\textwidth, keepaspectratio]{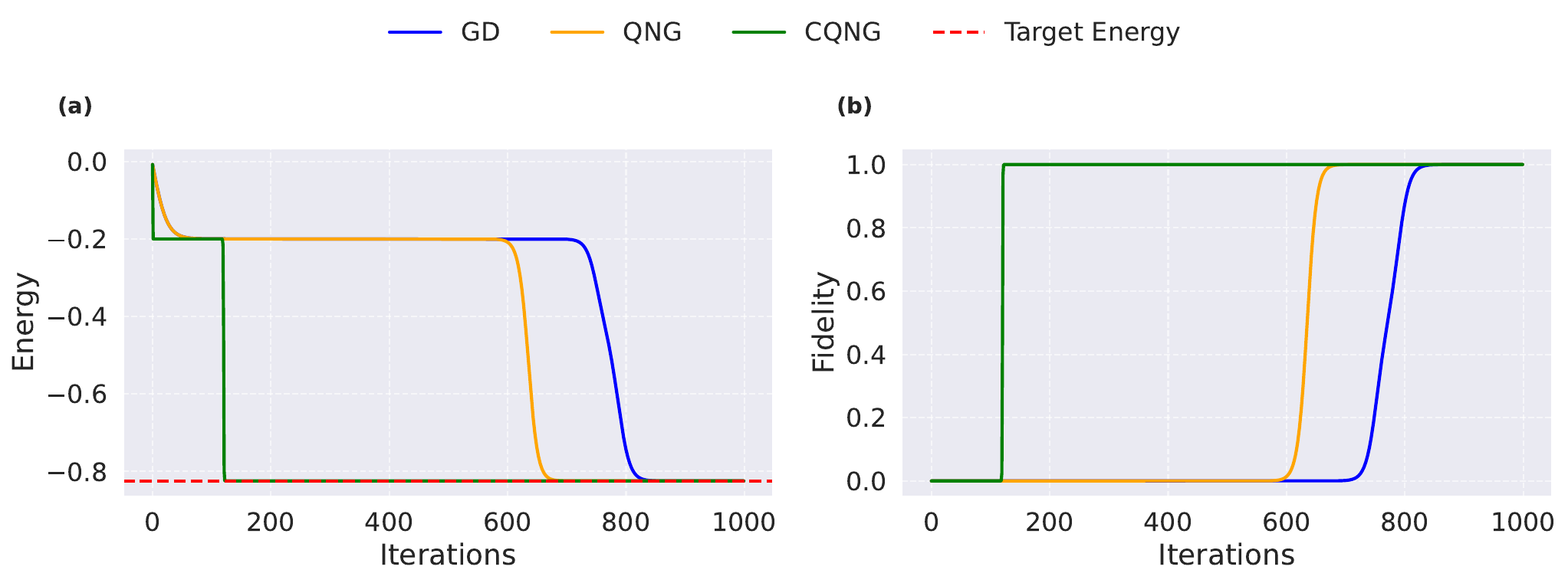}
    \caption{Same experimental conditions as in Fig.~\ref{fig:ex1_energy}, but with the initial point set to $[\pi/2,\pi/2,0]$. This setup is more challenging because the starting state lies adjacent to a nearly flat region of the cost landscape, causing optimizers to become trapped in local minima. As shown in the results, CQNG escapes these regions more quickly than the other methods.}
    \label{fig:ex1_energy_plateaus}
\end{figure}

\begin{figure}[H]  
    \centering
    \includegraphics[width=\textwidth, keepaspectratio]{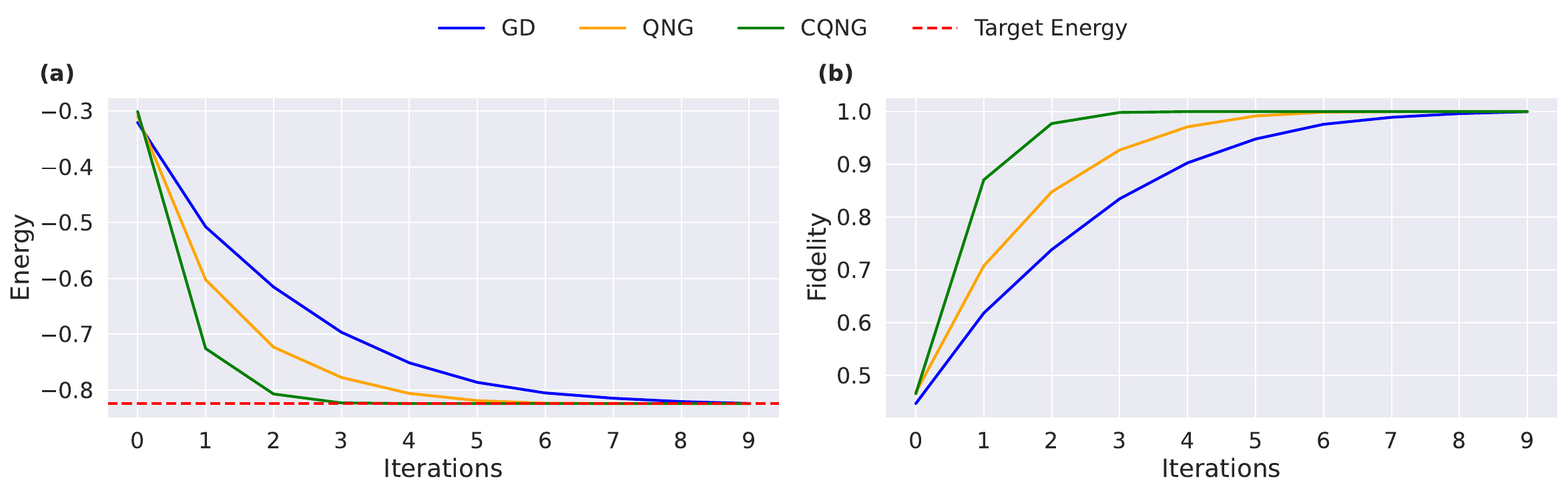}
    \caption{The average of 100 runs from different initial starting points. The hyperparameters, including the learning rate for GD and QNG, as well as the initial step size \( \alpha_0 \) used in \texttt{COBYLA} for CQNG, are tuned using a grid search.}
    \label{fig:ex1_energy_100}
\end{figure}

With this setup, Fig.~\ref{fig:ex1_energy}, Fig.~\ref{fig:ex1_energy_plateaus}, and Fig.~\ref{fig:ex1_energy_100} demonstrate that CQNG exhibits accelerated convergence compared to QNG and GD, reaching the target energy and high fidelity in fewer iterations. Furthermore, Fig.~\ref{fig:ex1_energy_plateaus} shows that CQNG escapes plateaus earlier.

\subsection{\textbf{Example 2: Heisenberg Model}}
\label{Example2}

In this example, we use the VQE to find the ground state of the Heisenberg Hamiltonian. 

The Heisenberg model is a fundamental quantum many-body system, widely studied for its rich physics and computational challenges. Its Hamiltonian for \(n\) spins is expressed as:
\begin{equation}
H = J \sum_{i=1}^{n-1} \left( \sigma_i^X \sigma_{i+1}^X + \sigma_i^Y \sigma_{i+1}^Y + \sigma_i^Z \sigma_{i+1}^Z \right) + h \sum_{i=1}^{n} \sigma_i^X,
\end{equation}
where \(\sigma_i^X, \sigma_i^Y, \sigma_i^Z\) are the Pauli operators acting on the \(i\)-th spin, \(J\) is the coupling constant, and \(h\) represents the strength of the external magnetic field. In this work, we consider the case where \(J = -1\) and \(h = -1\), making it a benchmark for testing quantum optimization methods.

\begin{figure}[H]
    \centering
    \includegraphics[width=0.5\textwidth]{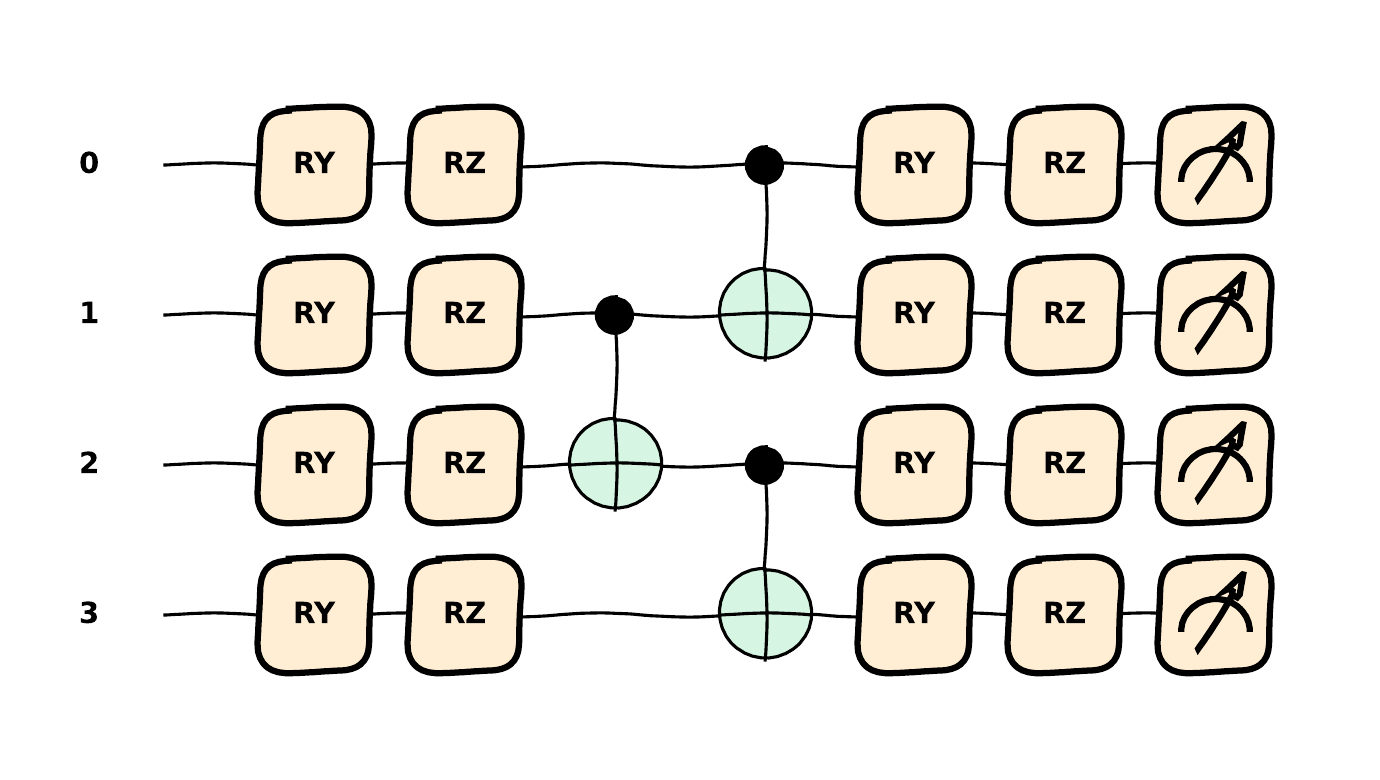}
    \caption{EfficientSU2 ansatz with two layers.}
    \label{fig:EfficientSU2_Ansatz}
\end{figure}

To approximate the ground state of \(H\), we employ the hardware-efficient \texttt{EfficientSU2} variational ansatz \cite{Kandala2017}. This ansatz combines layers of parameterized single-qubit \(R_Y\) and \(R_Z\) rotations with entangling CNOT gates, as illustrated in Fig.~\ref{fig:EfficientSU2_Ansatz}.

\begin{figure}[H]
    \centering
    \includegraphics[width=\textwidth]{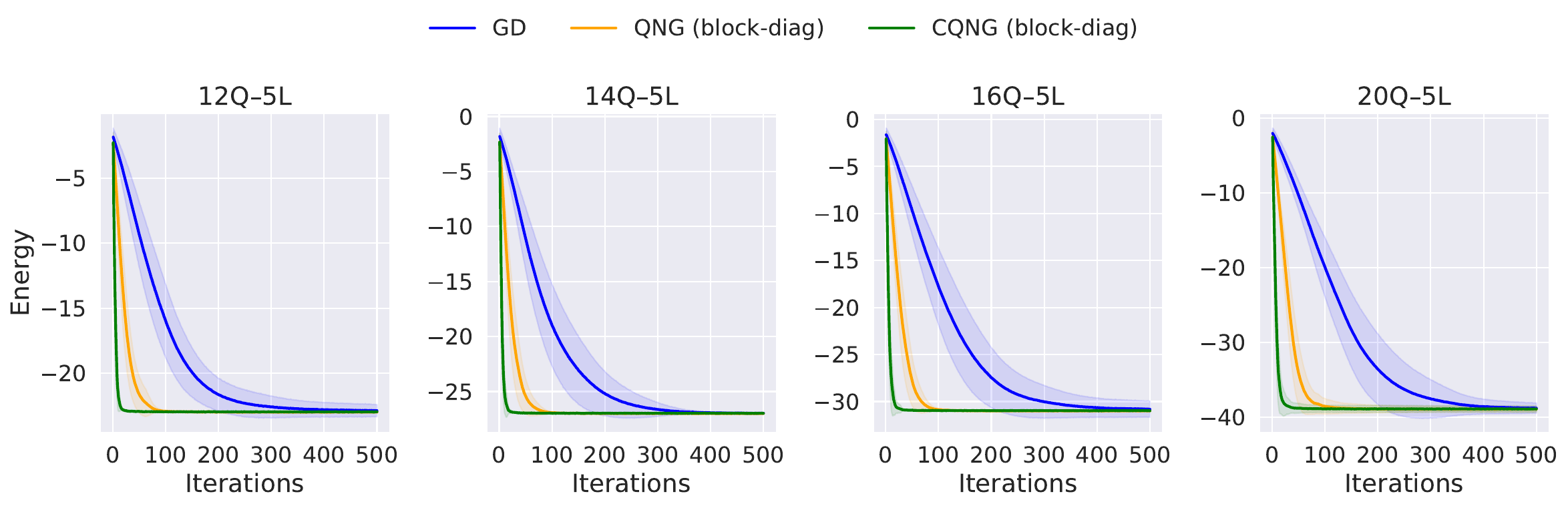}
    \caption{
    Average cost function value as a function of training iterations for GD, QNG, and CQNG, evaluated on 12, 14, 16, and 20 qubits with a fixed circuit depth of 5 layers—corresponding to 120, 140, 160, and 200 trainable parameters, respectively. Each data point represents an average over 30 runs with different initial parameters. The number of measurement shots is set to 10024, and a learning rate of 0.01 is used. Initial values $\beta_0 = 0.1$ and $\alpha_0 = 0.01$ are used for the \texttt{COBYLA} optimizer, which dynamically updates $\alpha_t$ and $\beta_t$ at each step according to Eq.~(\ref{eq:update_beta}).}
    \label{fig:ex2_energy_fix5L_fix_eta}
\end{figure}
\begin{figure}[H]  
    \centering
    \includegraphics[width=\textwidth, keepaspectratio]{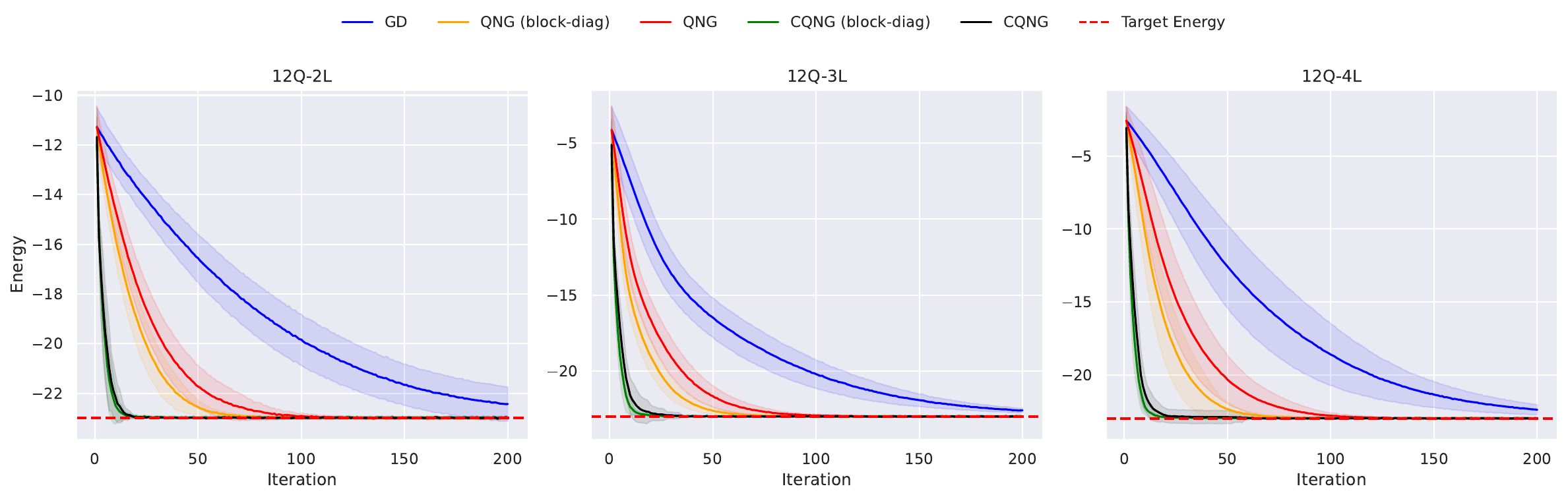}
    \caption{
    Same experimental conditions as in Fig.~\ref{fig:ex2_energy_fix5L_fix_eta}, but evaluated for circuit depths of 2, 3, and 4 with a fixed qubit count of 12—corresponding to 48, 72, and 96 trainable parameters, respectively.
    }
    \label{fig:ex2_energy_fix12Q_fix_eta}
\end{figure}
\begin{figure}[H]  
    \centering
    \includegraphics[width=\textwidth, keepaspectratio]{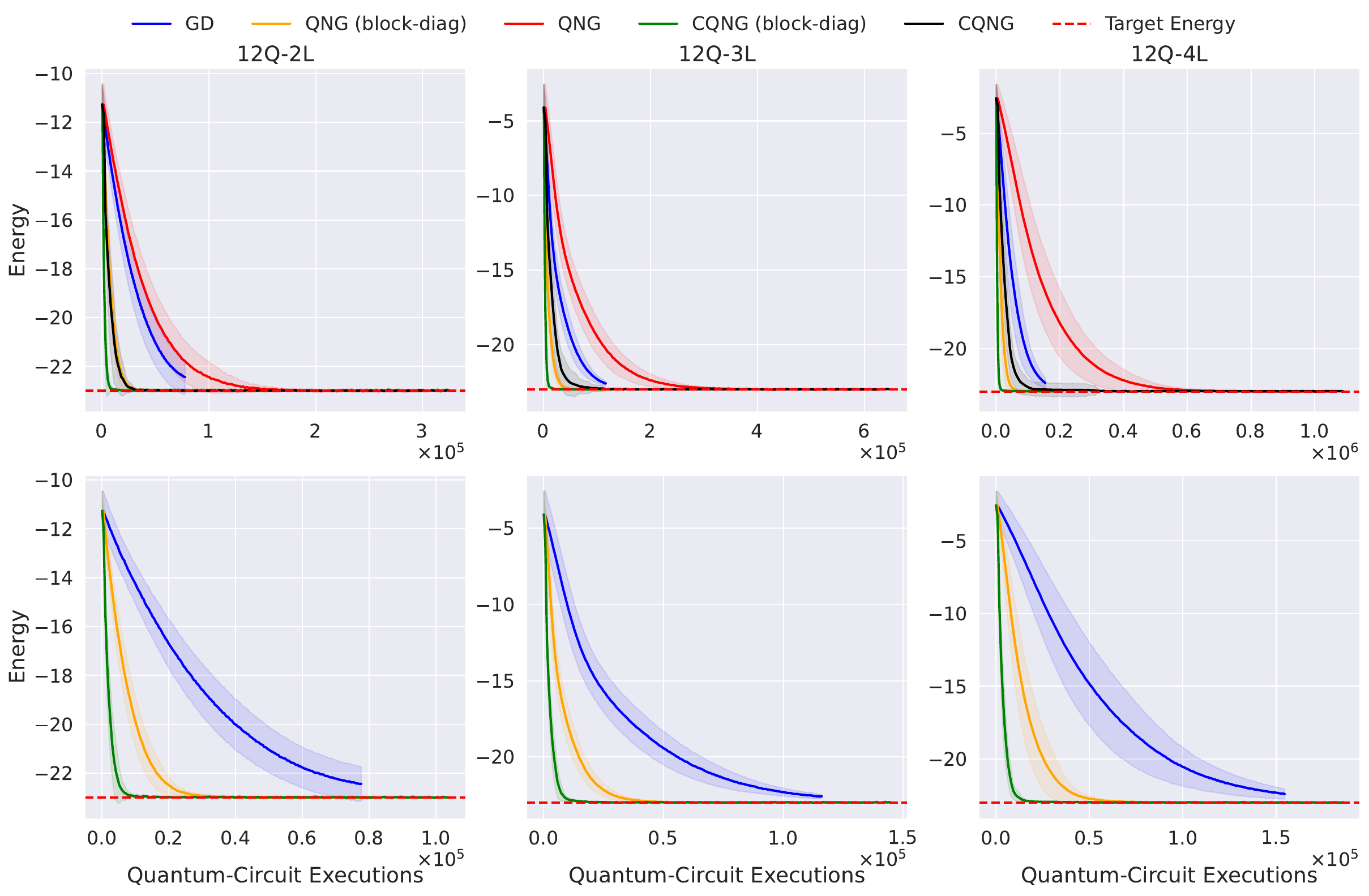}
    \caption{
    Same experiment as in Fig.~\ref{fig:ex2_energy_fix12Q_fix_eta}, but with energy plotted as a function of the number of quantum-circuit evaluations. The top row shows results for all optimizers: GD, QNG (block-diag), QNG, CQNG (block-diag), and CQNG. The bottom row presents the same data, but with QNGD and CQNG excluded for improved visual clarity.
    }
    \label{fig:circuits_Heisenberg}
\end{figure}

Like in Example 1, we benchmark the optimizers across different scenarios. In Figs.~\ref{fig:ex2_energy_fix5L_fix_eta} and \ref{fig:ex2_energy_fix12Q_fix_eta}, we fix the learning rate and plot the energy as a function of the number of iterations. In Fig.~\ref{fig:circuits_Heisenberg}, the energy is instead plotted as a function of the number of quantum-circuit evaluations. As the figures illustrate, CQNG consistently outperforms GD and QNGD in all scenarios by accelerating convergence and reducing the quantum resources required to reach lower energy values.

\subsection{\textbf{Example 3: Molecular Hamiltonian}}
\label{Molecular}

In this section, we use the Variational Quantum Eigensolver (VQE) algorithm to determine the ground-state energies (in Hartree) of molecular Hamiltonians. Specifically, we investigate three molecules: hydrogen four (H\textsubscript{4}), water (H\textsubscript{2}O), and carbon diatomic (C\textsubscript{2}), mapped onto systems of 8, 12, and 14 qubits, respectively. For all molecules, we employ a minimal STO-6G basis set and consider the neutral singlet state (charge = 0, multiplicity = 1). The Jordan–Wigner transformation is used to convert the fermionic Hamiltonians to corresponding qubit operators. The detailed molecular geometries are provided in Appendix~\ref{moleculargeometry}.
\begin{figure}[H]
  \centering
  \includegraphics[width=\linewidth]{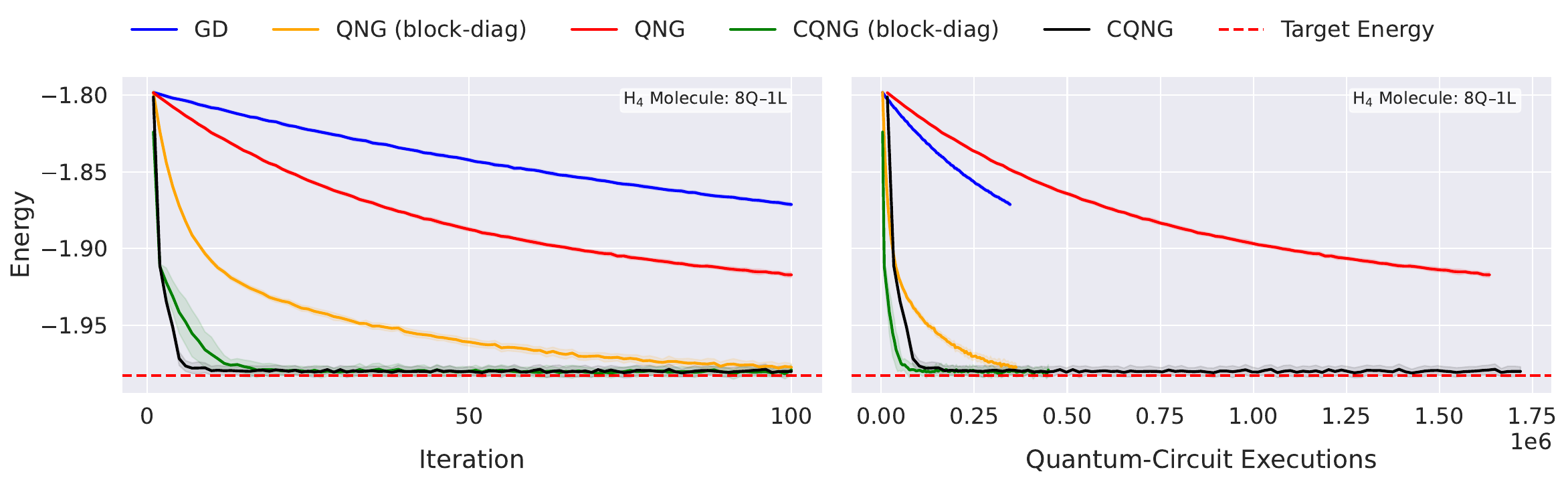}
  \caption{
  Average energy as a function of training iterations (left) and cumulative quantum-circuit executions (right) for GD, QNGD (block-diag), QNGD, CQNG (block-diag), and CQNG on an 8-qubit UCCSD ansatz with one repetition—corresponding to 26 trainable parameters. Each point represents the mean over 30 runs, all initialized at the Hartree–Fock reference state. Measurement shots = 50000 and learning rate = 0.01. Initial COBYLA parameters: \(\beta_0 = 0.1\), \(\alpha_0 = 0.01\).
  }
  \label{fig:H4_without_shots}
\end{figure}

The hydrogen four molecule (H\textsubscript{4}) is arranged in a square geometry with a side length of 2.25~Å. This configuration involves four active electrons distributed across four active spatial orbitals, resulting in an 8-qubit system after applying the Jordan–Wigner transformation. For the H\textsubscript{4} simulations, we utilize the Unitary Coupled-Cluster Singles and Doubles (UCCSD) ansatz~\cite{UCCSD}, initializing the quantum circuit in the Hartree–Fock reference state. It would be interesting to explore the multi-reference UCCSD (MR-UCCSD) ~\cite{MR-UCC} approach in future investigations.

For the water molecule (H\textsubscript{2}O), the tightly bound oxygen 1$s$ core orbital is frozen, leaving 8 active electrons distributed over 6 active spatial orbitals, corresponding to a 12-qubit system.
\begin{figure}[H]
    \centering
    \begin{minipage}[t]{0.2\textwidth} 
        \centering
        \makebox[0.3\textwidth]{(a)} \\ 
        \includegraphics[width=\textwidth]{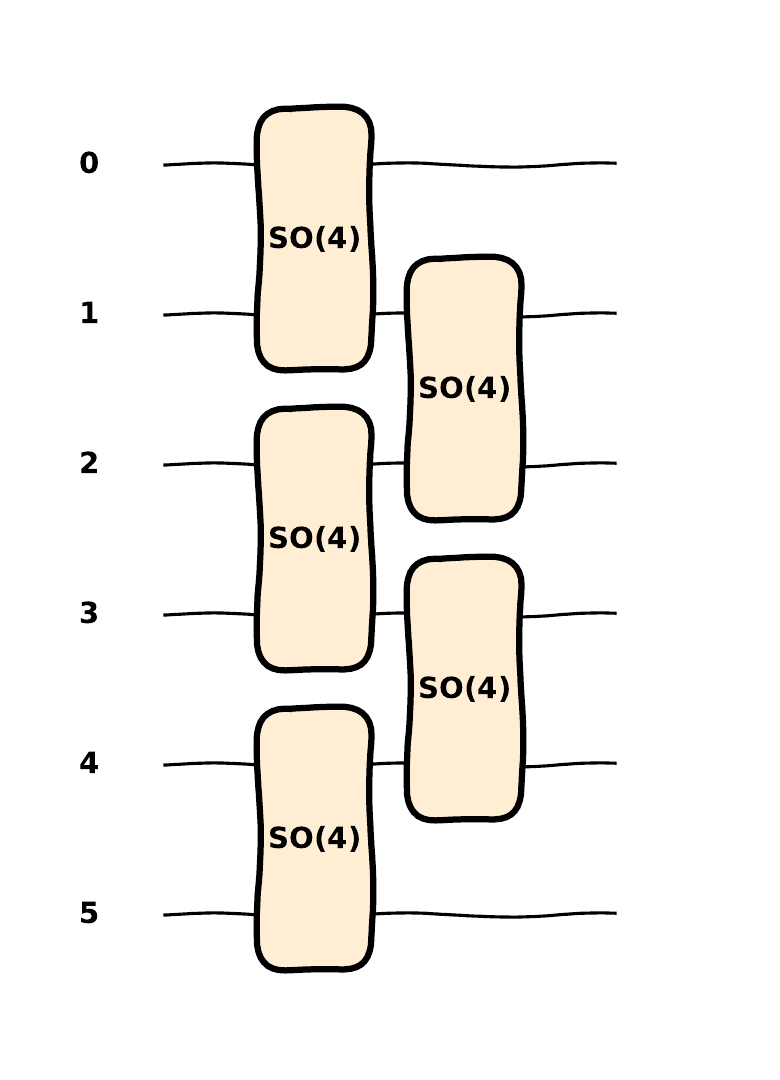} 
    \end{minipage}%
    \hfill
    \begin{minipage}[t]{0.65\textwidth} 
        \centering
        \makebox[0.65\textwidth]{(b)} \\ 
        \includegraphics[width=\textwidth]{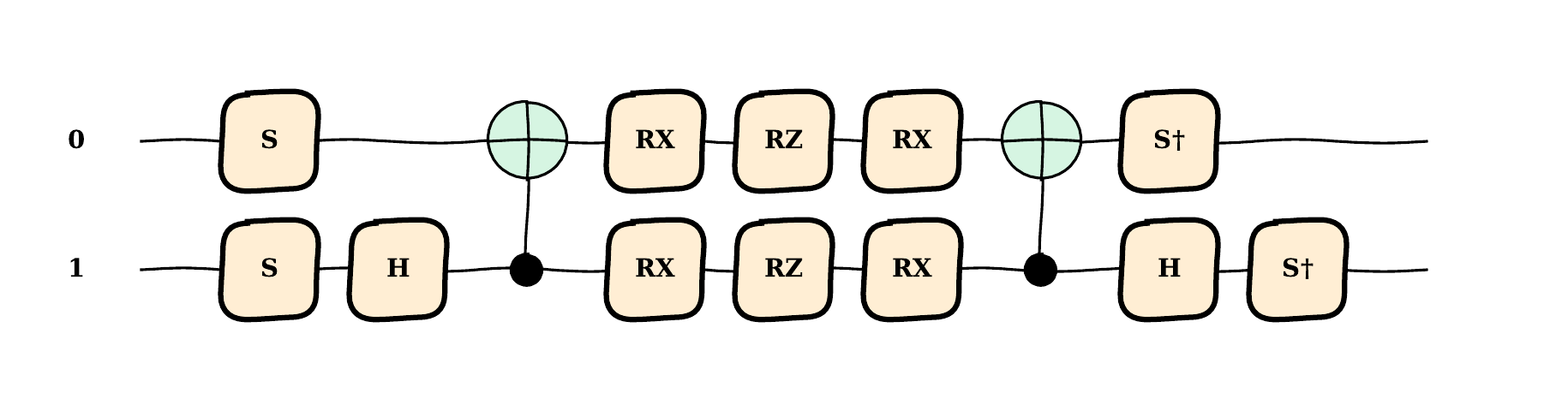} 
    \end{minipage}
    \caption{(a) The SO(4) ansatz applied to a 6-qubit system, shown with a single layer composed of universal SO(4) gates acting on adjacent qubit pairs. (b) The two-qubit \( \mathrm{SO}(4) \) gate is decomposed into a sequence of single-qubit operations—including phase gates \( S \) and \( S^\dagger \), Hadamard gates \( H \), and rotation gates \( R_X \) and \( R_Z \)—combined with entangling CNOT gates acting on qubit pairs.}

    \label{fig:SO4-ansatz}
\end{figure}
For the carbon diatomic molecule (C\textsubscript{2}), we freeze the tightly bound 1$s$ core orbitals of each carbon atom and the symmetric $\sigma_{g}$ bonding orbital formed from the carbon 2$s$ shells. This choice leaves 6 active electrons occupying 7 active spatial orbitals, mapped onto a 14-qubit system.

For the VQE simulations of H\textsubscript{2}O and C\textsubscript{2}, we employ the SO(4) ansatz (see Fig.~\ref{fig:SO4-ansatz}). This ansatz is NISQ-friendly and was originally developed in high-energy physics to describe strongly correlated systems~\cite{Yibin}. We begin by preparing the Hartree–Fock reference state $\ket{\mathrm{HF}}$ on the $n$-qubit register via single basis-state initialization. The SO(4) unitary gates are then applied across all qubit pairs, resulting in the variational circuit $U_{\mathrm{SO}(4)}(\boldsymbol{\theta}) \ket{\mathrm{HF}}$. The variational parameters $\boldsymbol{\theta}$ are initialized randomly within the interval $[-\pi/2, \pi/2]$.
\begin{figure}[H]
  \centering
  \includegraphics[width=\linewidth]{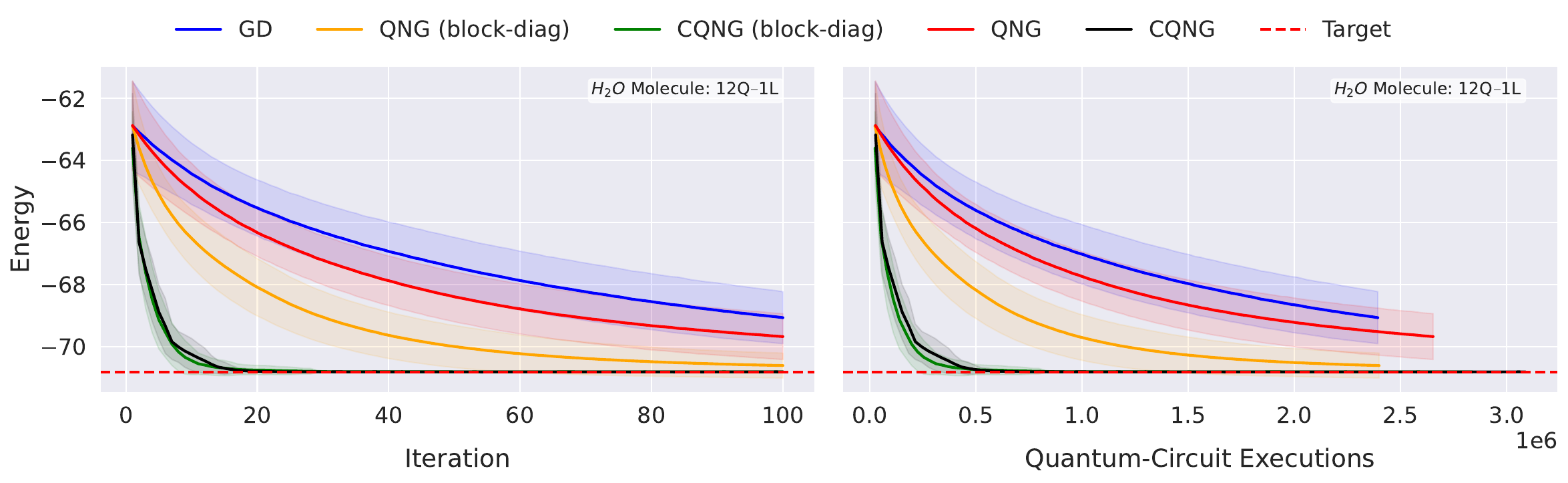}
  \caption{
  Average energy for the H\textsubscript{2}O molecule as a function of training iterations (left) and cumulative quantum-circuit executions (right), using GD, QNG (block-diag), QNG, CQNG (block-diag), and CQNG optimizers. The simulation employs a 12-qubit SO(4) ansatz with one layer—corresponding to 42 trainable parameters. Each data point represents an average over 30 runs with different initial parameters. The number of measurement shots is 10024, and the learning rate is set to 0.01. Initial COBYLA parameters: \(\beta_0 = 0.1\), \(\alpha_0 = 0.01\).
  }
  \label{fig:H2O_shots}
\end{figure}
\begin{figure}[H]
  \centering
  \includegraphics[width=\linewidth]{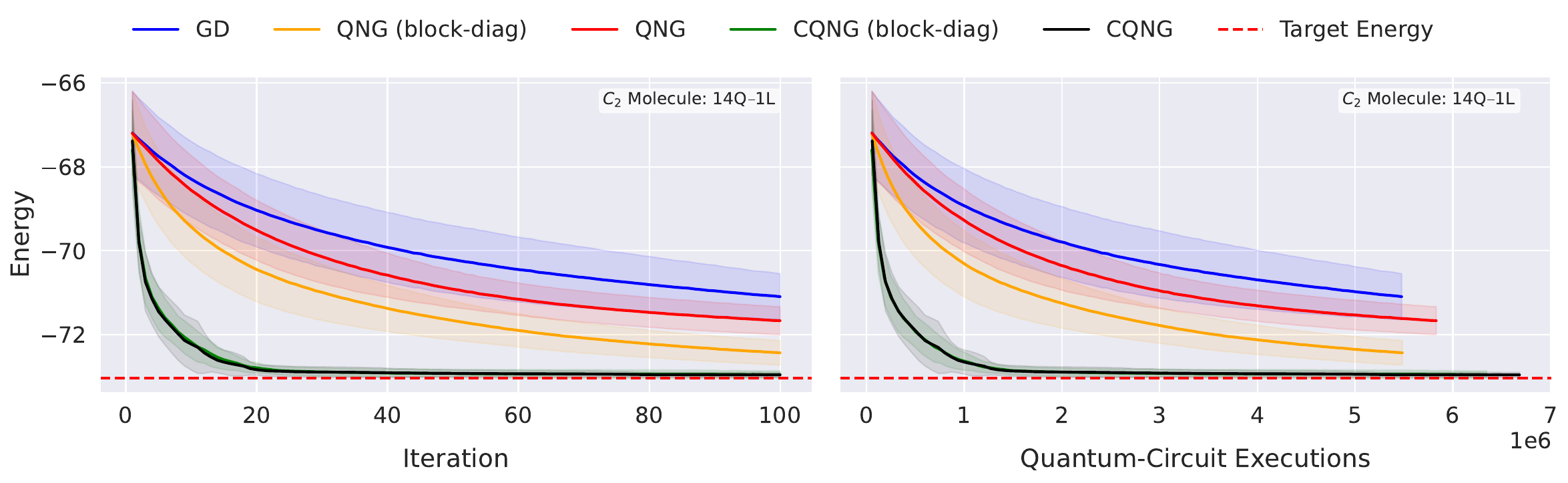}
  \caption{
  Average energy for the C\textsubscript{2} molecule as a function of training iterations (left) and cumulative quantum-circuit executions (right), using GD, QNGD (block-diag), QNGD, CQNG (block-diag), and CQNG optimizers. The simulation employs a 14-qubit SO(4) ansatz with one layer—corresponding to 78 trainable parameters. Each data point represents an average over 30 runs with different initial parameters. The number of measurement shots is 10024, and the learning rate is set to 0.01. Initial COBYLA parameters: \(\beta_0 = 0.1\), \(\alpha_0 = 0.01\).
  }
  \label{fig:C4_shots}
\end{figure}
\begin{figure}[H]
  \centering
  \includegraphics[width=\linewidth]{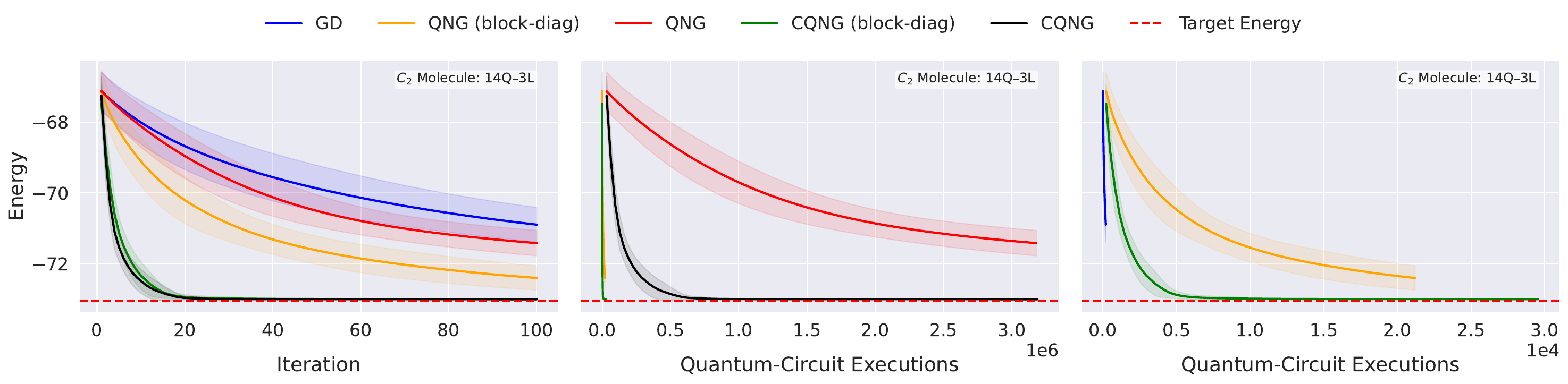}
  \caption{
  Same experimental conditions as in Fig.~\ref{fig:C4_shots}, but with three layers in the ansatz. For simplicity, no measurement-shot noise is included. The plots show the average energy for the C\textsubscript{2} molecule as a function of training iterations (left) and cumulative quantum-circuit executions (middle and right). The right panel is identical to the middle one but omits QNGD and CQNG to improve visual clarity when comparing QNGD (block-diag) and CQNG (block-diag). In this setup, the circuit includes 234 trainable parameters.
  }
  \label{fig:C4_3R}
\end{figure}
In this experiment, as in previous ones, CQNG converged to the target ground-state energies in fewer iterations and with fewer quantum-circuit executions than QNGD—particularly for the larger H\textsubscript{2}O and C\textsubscript{2} systems—highlighting the benefits of integrating conjugate-gradient updates into the natural-gradient framework for quantum chemistry simulations.
\section{Conclusions and Outlook}
\label{outlook}
We have introduced the \textit{Modified Conjugate Quantum Natural Gradient (CQNG)}, an optimization algorithm that extends the \textit{Quantum Natural Gradient (QNG)} by incorporating principles from the nonlinear conjugate gradient method. Unlike conventional QNG, which relies on a fixed learning rate, CQNG dynamically optimizes both the \textit{step size} (\(\alpha_t\)) and the \textit{conjugate coefficient} (\(\beta_t\)) at each iteration. 

Numerical simulations demonstrate that this adaptive approach significantly accelerates convergence and reduces quantum resource requirements compared to standard QNG, making CQNG a promising optimization method for variational quantum algorithms. However, the conjugacy condition, \( \bm{d}_t^T \bm{H} \bm{d}_{t-1} = 0 \), where \( \bm{H} \) denotes the Hessian of the cost function, may not be strictly satisfied at every iteration of the optimization process. This motivates the term \textit{"Modified Conjugate"} to reflect the practical adjustments made to improve performance while retaining the core advantages of conjugate gradient methods.

The Quantum Natural Gradient (QNG) algorithm, which uses the Fubini-Study metric, requires \( O(m^2) \) quantum-circuit evaluations per iteration, where \( m \) is the number of variational parameters. Our conjugate version (CQNG) has the same \(O(m^{2})\) scaling, as it also constructs the metric, and adds only a constant, \(m\)-independent overhead to solve the two-parameter \((\alpha,\beta)\) subproblem in Eq.~\eqref{eq:update_beta}. This extra \(O(1)\) work is negligible compared to the \(O(m^{2})\) metric cost. Since both QNG and CQNG account for the curvature of the parameter space, CQNG follows this curvature more closely by additionally incorporating the previous search direction. As a result, it typically requires fewer iterations to converge, leading to a lower total number of quantum-circuit evaluations compared to standard QNG.

CQNG can be further developed in several directions. Similar to QNG, it can be adapted for noisy and nonunitary circuits, following the approaches in \cite{Koczor}. The use of simultaneous perturbation stochastic approximation for the quantum Fisher information \cite{Gacon}, randomness-based methods \cite{Kolotouros}, as well as Stein's identity~\cite{Halla2}, represents a promising avenue for our future research. Moreover, the methods introduced in CQNG could also be explored for applications in time-dependent quantum optimization problems, potentially extending its applicability to broader areas of quantum computing.

\section*{Acknowledgements}
This work was supported by the Ministry of Science, Research, and Culture of the State of Brandenburg through the Centre for Quantum Technologies and Applications (CQTA) at DESY (Germany) and by the German Ministry of Education and Research (BMBF) under project NiQ (Grant No.\ 13N16203). The author thanks Davide Materia for valuable discussions on quantum chemistry.

\begin{figure}[H]
\centering
\includegraphics[width=0.2\textwidth]{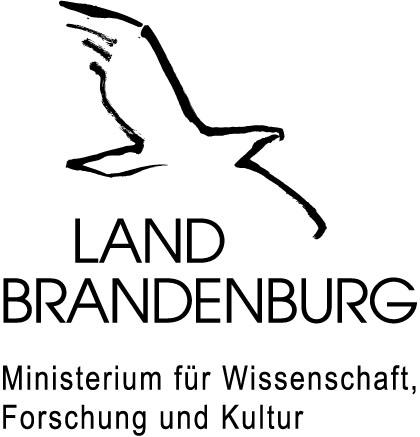}
\end{figure}

\appendix
\section{The molecular geometry}
\label{moleculargeometry}
In this section, we provide the molecular geometries in \texttt{xyz-format} (atomic units) used in Section~\ref{Molecular}.

\begin{itemize}
\item Hydrogen Four: H\textsubscript{4}~\cite{qBang}:
\begin{verbatim}
4
*
H  2.1213  2.1213  0.0
H  2.1213 -2.1213  0.0
H -2.1213  2.1213  0.0
H -2.1213 -2.1213  0.0
\end{verbatim}

\item Water molecule: H\textsubscript{2}O~\cite{qBang}: 

\begin{verbatim}
3
*
O  0.0000  0.0000  0.0000
H  0.8081  1.0437  0.0000
H  0.8081 -1.0437  0.0000
\end{verbatim}

\item Carbon diatomic: C\textsubscript{2}~\cite{NISTCCCDB2022}: 

\begin{verbatim}
2
*
C  0.0000  0.0000  1.1654
C  0.0000  0.0000 -1.1654
\end{verbatim}

\end{itemize}

\section{Additional Experiments and Figures}
In this supplementary experimental section, we use the same Hamiltonian and ansatz as in Example~\ref{Example2}. 
\subsection{Hyperparameter Tuning}
\begin{figure}[H]  
    \centering
    \includegraphics[width=\textwidth, keepaspectratio]{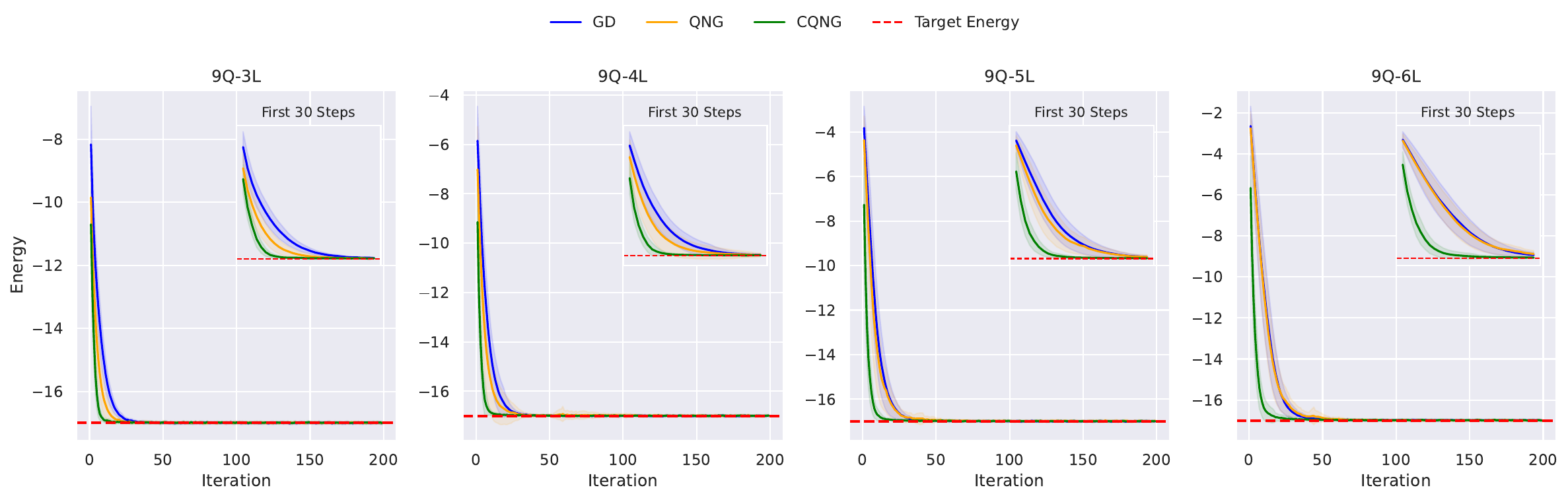}
    \caption{
    Average cost function value as a function of training iterations for GD, QNGD, and CQNG, evaluated at circuit depths of 3, 4, 5, and 6 with a fixed qubit count of 9—corresponding to 54, 72, 90, and 108 trainable parameters, respectively. Each data point represents an average over 30 runs with different initial parameters. The number of measurement shots is set to 10024. The learning rates for GD and QNGD, as well as the initial step size $\alpha_0$ in CQNG, are optimized through hyperparameter tuning \cite{Optuna}.
    }
    \label{fig:ex2_optuna_energy_fix9Q}
\end{figure}

In Fig.~\ref{fig:ex2_optuna_energy_fix9Q}, we determine the optimal learning rates for GD and QNGD, as well as the initial step size \( \alpha_0 \) for \texttt{COBYLA}, through hyperparameter tuning.
\subsection{Sensitivity of the Initial Parameters \(\alpha_0\) and \(\beta_0\)}

\begin{figure}[H]
  \centering
  \includegraphics[width=\linewidth]{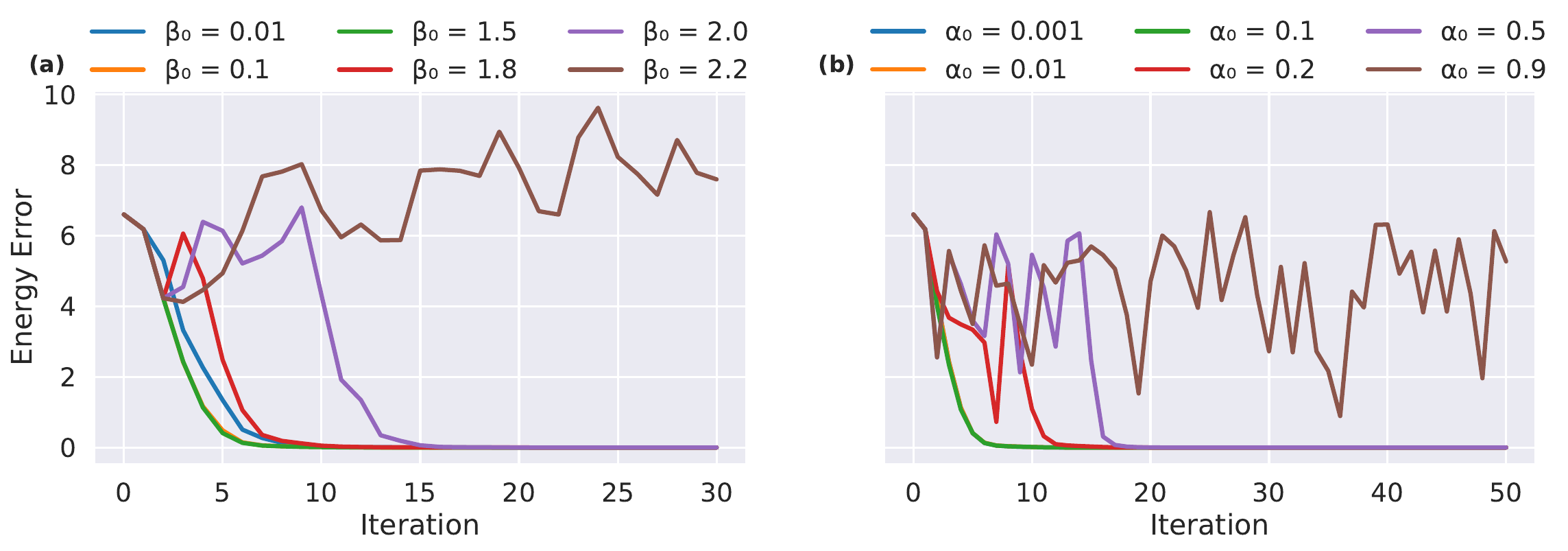}
  \caption{Sensitivity of CQNG to the initial hyperparameters.  
  (\textbf{a}) Energy error $\Delta E$ vs.\ iteration for various $\beta_0$ values with fixed $\alpha_0 = 0.1$, where $\Delta E = |E(\theta_t) - E_\text{ref}|$ denotes the absolute error with respect to the reference ground-state energy.
  (\textbf{b}) Energy error vs.\ iteration for various $\alpha_0$ values with fixed $\beta_0 = 0.1$.}
  \label{fig:par_sensitivity}
\end{figure}

We evaluate the effect of the hyperparameters \(\alpha_0\) and \(\beta_0\) on convergence using a 4-qubit, 3-layer ansatz with an exact (shot-noise-free) simulator. Figure~\ref{fig:par_sensitivity}(a) shows energy-error curves for fixed \(\alpha_0 = 0.1\) as \(\beta_0\) ranges from 0.1 to 2.2; convergence remains smooth for \(\beta_0 \lesssim 1.8\), but higher values induce oscillations and eventual divergence. Conversely, Fig.~\ref{fig:par_sensitivity}(b) fixes \(\beta_0 = 0.1\) and varies \(\alpha_0\) from 0.001 to 0.9; here, \(\alpha_0 \lesssim 0.2\) yields stable decay, while larger \(\alpha_0\) leads to failure. In practice, setting \(\alpha_0\) equal to the CQNG step size and \(\beta_0 = 0.1\) offers robust, rapid convergence across our test cases. These results confirm that the method is only weakly sensitive to \(\alpha_0\) and \(\beta_0\) within reasonable bounds, but can diverge if either parameter is set too large.

\newpage

\begin{thebibliography}{99}
\bibitem{Cerezo2021}
M. Cerezo, A. Arrasmith, R. Babbush, S. C. Benjamin, S. Endo, K. Fujii, et al., ``Variational Quantum Algorithms,'' Nature Reviews Physics, 3, 625-644 (2021). \url{https://doi.org/10.1038/s42254-021-00348-9}.

\bibitem{McClean2016}
J. R. McClean, J. Romero, R. Babbush and A. Aspuru-Guzik, ``The theory of variational hybrid quantum-classical algorithms,'' New Journal of Physics, 18, 023023 (2016). \url{https://doi.org/10.1088/1367-2630/18/2/023023}.

\bibitem{Bharti2022}
K. Bharti, A. Cervera-Lierta, T. H. Kyaw, T. Haug, S. Alperin-Lea, A. Anand, et al., ``Noisy intermediate-scale quantum algorithms,'' Reviews of Modern Physics, 94, 015004 (2022). \url{https://doi.org/10.1103/RevModPhys.94.015004}.

\bibitem{Peruzzo2014}
A. Peruzzo, J. McClean, P. Shadbolt, M.-H. Yung, X.-Q. Zhou, P. J. Love, A. Aspuru-Guzik, and J. L. O'Brien, ``A variational eigenvalue solver on a photonic quantum processor,'' Nature Communications 5, 4213 (2014). \url{https://www.nature.com/articles/ncomms5213}{doi:10.1038/ncomms5213}.


\bibitem{Stokes2020}
J. Stokes, J. Izaac, N. Killoran, and G. Carleo, ``Quantum Natural Gradient,'' Quantum 4, 269 (2020). \url{https://doi.org/10.22331/q-2020-05-25-269}.

\bibitem{Meyer}
J. Jakob Meyer,, ``Fisher Information in Noisy Intermediate-Scale Quantum Applications
,'' Quantum 5, 539 (2021). \url{https://doi.org/10.22331/q-2021-09-09-539}.

\bibitem{Amari1998}
S.-I. Amari, ``Natural Gradient Works Efficiently in Learning,'' Neural Computation 10 (2), 251-276 (1998). \url{https://doi.org/10.1162/089976698300017746}

\bibitem{Katabarwa2022}
A. Katabarwa, S. Sim, D. E. Koh, and P.-L. Dallaire-Demers, ``Connecting geometry and performance of two-qubit parameterized quantum circuits,'' \textit{Quantum} 6, 782 (2022). \url{https://doi.org/10.22331/q-2022-08-23-782}.

\bibitem{Koczor}
B. Koczor and  C. Benjami, ``Quantum natural gradient generalized to noisy and nonunitary circuits,'' \textit{Phys. Rev. A} \textbf{106}, 062416 (2022). \url{https://doi.org/10.1103/PhysRevA.106.062416}.

\bibitem{Gacon}
J. Gacon, C. Zoufal, G. Carleo and S. Woerner, ``Simultaneous Perturbation Stochastic Approximation of the Quantum Fisher Information
,'' Quantum 5, 567 (2021). \url{https://doi.org/10.22331/q-2021-10-20-567}.

\bibitem{Kolotouros}
I. Kolotouros and P. Wallden, ``Random Natural Gradient
,'' Quantum 8, 1503 (2024). \url{https://doi.org/10.22331/q-2024-10-22-1503}.

\bibitem{Halla}
M. Halla,, ``Quantum Natural Gradient with Geodesic Corrections for Small Shallow Quantum Circuits,'' Phys. Scr. 100, 055121 (2025). \url{https://doi.org/10.1088/1402-4896/add05e}.

\bibitem{Halla2}
M. Halla,, ``Estimation of Quantum Fisher Information via Stein's Identity in Variational Quantum Algorithms,'' arXiv:2502.17231 (2025). \url{https://doi.org/10.48550/arXiv.2502.17231}.

\bibitem{Hestenes1952}
M. R. Hestenes and E. Stiefel, ``Methods of Conjugate Gradients for Solving Linear Systems,'' Journal of Research of the National Bureau of Standards, 49(6), 409-436 (1952). \url{https://doi.org/10.6028/jres.049.044}.

\bibitem{Fletcher1964}
R. Fletcher and C. M. Reeves, ``Function Minimization by Conjugate Gradients,'' The Computer Journal, 7(2), 149-154 (1964). \url{https://doi.org/10.1093/comjnl/7.2.149}.

\bibitem{Saad2003}
Y. Saad, ``Iterative Methods for Sparse Linear Systems,'' 2nd ed., SIAM, Philadelphia, 2003. \url{https://doi.org/10.1137/1.9780898718003}.


\bibitem{Pascanu2013}
R. Pascanu and Y. Bengio, ``Revisiting Natural Gradient for Deep Networks,'' arXiv preprint arXiv:1301.3584 (2013). \url{https://arxiv.org/abs/1301.3584}.

\bibitem{Andrei2020}
N. Andrei, \textit{Nonlinear Conjugate Gradient Methods for Unconstrained Optimization}, Springer, 2020. 
\url{https://doi.org/10.1007/978-3-030-42950-8}.

\bibitem{Polak1969}
E. Polak and G. Ribi\`ere, ``Note sur la convergence de m\'ethodes de directions conjugu\'ees,'' 
\textit{ESAIM Math. Model. Numer. Anal.}, vol. 3, pp. 35--43, 1969.

\bibitem{Fletcher1964}
R. Fletcher and C. M. Reeves, ``Function minimization by conjugate gradients,'' 
\textit{Comput. J.}, vol. 7, pp. 149--154, 1964.

\bibitem{Dai1999}
Y.-H. Dai and Y. Yuan, ``A nonlinear conjugate gradient method with a strong global convergence property,'' 
\textit{SIAM J. Optim.}, vol. 10, pp. 177--182, 1999.

\bibitem{McClean2018}
J. R. McClean, S. Boixo, V. N. Smelyanskiy, R. Babbush, and H. Neven, ``Barren plateaus in quantum neural network training landscapes,'' Nat. Commun. 9, 4812 (2018). \url{https://doi.org/10.1038/s41467-018-07090-4}.
\bibitem{Patel2025}
S.~Patel, P.~Jayakumar, T.~C.~Yen, and A.~F.~Izmaylov,
``Quantum Measurement for Quantum Chemistry on a Quantum Computer,''
arXiv:2501.14968 (2025).
\url{https://arxiv.org/abs/2501.14968}.
\bibitem{Gonthier2022}
J.~F.~Gonthier, M.~D.~Radin, C.~Buda, E.~J.~Doskocil, C.~M.~Abuan, and J.~Romero,
``Measurements as a roadblock to near-term practical quantum advantage in chemistry: Resource analysis,''
\textit{Phys. Rev. Research} \textbf{4}, 033154 (2022).
\url{https://doi.org/10.1103/PhysRevResearch.4.033154}.


\bibitem{qBang}
D. Fitzek, R. S. Jonsson, W. Dobrautz, and C. Schäfer, “Optimizing Variational Quantum Algorithms with qBang: Efficiently Interweaving Metric and Momentum to Navigate Flat Energy Landscapes,” Quantum 8, 1313 (2024). \url{https://doi.org/10.22331/q-2024-04-09-1313}

\bibitem{UCCSD}
P. K. Barkoutsos, J. F. Gonthier, I. Sokolov, N. Moll, G. Salis, A. Fuhrer, M. Ganzhorn, D. J. Egger, M. Troyer, A. Mezzacapo, S. Filipp, and I. Tavernelli, “Quantum algorithms for electronic structure calculations: Particle-hole Hamiltonian and optimized wave-function expansions,” Phys. Rev. A 98, 022322 (2018). \url{https://doi.org/10.1103/PhysRevA.98.022322}

\bibitem{MR-UCC}
Di Wu, C.L. Bai, H. Sagawa, H.Q. Zhang, “Multi-Reference UCCSD Variational Quantum Algorithm for Molecular Ground State Energies,” arXiv:2408.16523 (2025). \url{https://doi.org/10.48550/arXiv.2408.16523}

\bibitem{Yibin}
Y. Guo , T. Angelides , K. Jansen and S. Kuehn, ``Concurrent VQE for Simulating Excited States of the Schwinger Model,'' (2024). \url{https://doi.org/10.48550/arXiv.2407.15629}.

\bibitem{NISTCCCDB2022}
R. D. Johnson III, ed., “NIST Computational Chemistry Comparison and Benchmark Database, NIST Standard Reference Database Number 101, Release 22, May 2022,” National Institute of Standards and Technology. \url{http://cccbdb.nist.gov/}. \url{https://doi.org/10.18434/T47C7Z}

\bibitem{Schuld}
M. Schuld, V. Bergholm, C. Gogolin, J. Izaac, and N. Killoran, ``Evaluating analytic gradients on quantum hardware,'' Phys. Rev. A 99, 032331 (2019). \url{https://doi.org/10.1103/PhysRevA.99.032331}

\bibitem{Mari}
A. Mari, T. R. Bromley, and N. Killoran, ``Estimating the gradient and higher-order derivatives on quantum hardware,'' Phys. Rev. A 103, 012405 (2021). \url{https://doi.org/10.1103/PhysRevA.103.012405}

\bibitem{Wierichs}
D. Wierichs, J. Izaac, C. Wang, and C. Yen-Yu Lin, ``General parameter-shift rules for quantum gradients,'' Quantum 6, 677 (2022). \url{https://doi.org/10.22331/q-2022-03-30-677}.
\bibitem{PennyLane}
V. Bergholm, J. Izaac, M. Schuld, C. Gogolin, S. Ahmed, V.
Ajith, M. S. Alam, G. Alonso-Linaje, B. AkashNarayanan, A.
Asadi et al., ``Pennylane: Automatic differentiation of hybrid
quantum-classical computations,'' \url{https://doi.org/10.48550/arXiv.1811.04968}

\bibitem{Optuna}
T. Akiba, S. Sano, T. Yanase, T. Ohta, and M. Koyama, ``Optuna: A Next-generation Hyperparameter Optimization Framework,'' 2019. \url{https://optuna.org}.

\bibitem{Kandala2017}
A. Kandala, A. Mezzacapo, K. Temme, M. Takita, M. Brink, J. M. Chow, and J. M. Gambetta, ``Hardware-efficient variational quantum eigensolver for small molecules and quantum magnets,'' Nature 549, 242-246 (2017). \url{https://doi.org/10.1038/nature23879}.



\end{thebibliography}
\end{document}